\documentclass[showpacs,prl,amsmath,draft,11pt]{revtex4}
\usepackage{xspace}

\newcommand{\etal}{{\it et~al.}\xspace}
\newcommand{\ie}{{\it i.e.}\xspace}
\newcommand{\ave}[1]{\left\langle#1 \right\rangle}
\newcommand{\abstractCite}[1]{[#1]}

\begin{document}
\title{Reply to ``Comment on `Self-organized Criticality and
Absorbing States: Lessons from the Ising
Model'{''}}
\date{\today}
\author{Gunnar Pruessner}
\email{g.prussner@imperial.ac.uk}
\homepage{http://www.ma.imperial.ac.uk/~pruess}
\affiliation{Department of Mathematics,
Imperial College London,
180 Queen's Gate,
London SW7 2AZ,
UK}

\author{Ole Peters}
\email{ole@santafe.edu}
\homepage{http://www.santafe.edu/~ole/}
\affiliation{Department of Atmospheric Sciences,
University of California, Los Angeles,
405 Hilgard Avenue,
Los Angeles,
California 90095-1565,
USA}

\begin{abstract}
In \abstractCite{\bibinfo{journal}{Braz. J. Phys.}
\textbf{\bibinfo{volume}{30}}, \bibinfo{pages}{27}
(\bibinfo{year}{2000})} Dickman \etal suggested that self-organized
criticality can be produced by coupling the activity of an
absorbing state model to a dissipation mechanism and adding an
external drive. We analyzed the proposed mechanism in
\abstractCite{\bibinfo{journal}{Phys.~Rev.~E}
\textbf{\bibinfo{volume}{73}},
  \bibinfo{pages}{025106(R)} (\bibinfo{year}{2006})}
and found that if this mechanism is at
work, the finite-size scaling found in self-organized criticality 
will depend on the details
of the implementation of dissipation and driving.
In the preceding comment 
\abstractCite{\bibinfo{journal}{Phys.~Rev.~E}
\textbf{\bibinfo{volume}{XX}},
  \bibinfo{pages}{XXXX} (\bibinfo{year}{2008})}, 
  Alava \etal show that one avalanche exponent in
the AS approach becomes independent of dissipation 
and driving. In our reply we clarify their findings 
and put them in the context of the original article.
\end{abstract}

\pacs{
05.65.+b, 
05.50.+q 
05.70.Jk, 
64.60.Ht 
}
\maketitle
\newpage

In \cite{PetersPruessner:2006} we discussed the implications of the
absorbing state (AS) mechanism \cite{DickmanETAL:2000} 
if it were to underlie
self-organized criticality (SOC). One of the key ingredients of the
AS-mechanism is that dissipation and driving are made to 
vanish in the thermodynamic
limit. 
We showed that criticality would be reached in the 
thermodynamic limit for almost any choice of the scaling 
with system size of dissipation and driving 
(with the effective temperature vanishing, contrary to what is stated in
\cite{AlavaETAL:2008}). While this choice is thus not important
to answer the question {\em if} the critical point will be reached, 
it is important when addressing the question {\em how} it 
will be reached.
In particular, we showed that the critical point 
would be reached in the limit of slow
drive $\omega-\kappa>\beta/\nu$ (``too fast'' in
\cite{PetersPruessner:2006}), but that the 
observed
finite-size scaling 
exponents
would depend on $\omega-\kappa$, and the relative correlation 
length $\xi/L$ would vanish asymptotically.

The present discussion can be phrased in terms of two statements. 
1) SOC is universal. 2) The AS-mechanism is solely responsible 
for SOC. In \cite{PetersPruessner:2006} we showed that these are
mutually exclusive. If the AS-mechanism is the reason for SOC, 
then SOC cannot be universal. This would weaken 
SOC considerably since studying models as simple as sandpiles 
is only sensible if these systems show universal behavior. The 
other possibility is that SOC is universal, which would weaken 
the AS-mechanism, because it does not predict universality. 
Our analysis was restricted to the finite-size scaling of AS 
observables, such as the order parameter, the correlation 
length, and the susceptibility, whereas the preceding comment 
\cite{AlavaETAL:2008} by Alava \etal refers to avalanche 
characteristics. In the following we assume that what we 
found out about the universality of AS observables also 
applies to avalanches. We emphasize that we do not know 
whether this is true; this assumption is made in order 
to be able to reply to the comment, which makes the same 
assumption. 

In the following we distinguish between the AS approach intended to
explain SOC, and SOC itself. Alava \etal do not make this distinction
explicit. In the AS approach driving and dissipation rates are tuned
(bulk dissipation as $L^{-\kappa}$ and driving as $L^{-\omega}$),
whereas in SOC (boundary dissipation, driving on a separate time scale)
they are set implicitly by the dynamics of the system. If the AS
approach applies to SOC, then SOC behaviour is obtained within the AS
approach by taking the thermodynamic limit. Both, our original article
\cite{PetersPruessner:2006} as well as the preceding comment
\cite{AlavaETAL:2008}, are concerned solely with the characterisation of
the AS approach.

It is important to stress that in \cite{PetersPruessner:2006} we did \emph{not}
claim that any exponents, neither those describing avalanches nor those 
characterising the activity, in standard SOC models 
are non-universal. This deserves clarification, because the opposite is stated 
in the abstract of \cite{AlavaETAL:2008}.
As discussed in \cite{PetersPruessner:2006}, there are instances of SOC
exponents being identical across a wide range of models
\cite{PanETAL:2005,ChristensenETAL:2004,Malthe-Sorenssen:1999}, while others are not
\cite{ChristensenETAL:2004,Malthe-Sorenssen:1999}.

We do claim (with the proviso stated above),
however, that avalanche size exponents would be non-universal if the AS
mechanism \cite{DickmanETAL:2000} 
was applicable to SOC 
\cite{PetersPruessner:2006}.
Alava \etal challenge this claim by showing that \emph{within the 
AS approach} $\tau_s$ is independent of the scaling of 
driving and dissipation in the slow driving limit.
This limit corresponds to the separation of time scales in SOC.

In \cite{PetersPruessner:2006} we explicitly mention avalanche exponents
only once: ``In the AS approach
also the avalanche size exponents show a clear, immediate dependence on
the choice of the two exponents $\kappa$ and $\omega$.''
It is correct
that the avalanche size exponents $\tau_s$ and $D_L$ (see below) depend
on $\kappa$ and $\omega$, but as Alava \etal show, for $\omega$ large
enough $\tau_s$ becomes independent of $\kappa$ and $\omega$. In this
case, as we show below, $D_L$ nonetheless depends on $\kappa$. 
The non-universality of $D_L$ within the AS approach implies that 
it does not explain universal SOC. This is the same conclusion we 
reached by studying AS observables (order parameter, correlation 
length, susceptibility).

We agree
that the derivation in \cite{PetersPruessner:2006} necessarily breaks
down in the large $\omega$ limit. We explicitly assumed finite bounds
for both, $\kappa$ and $\omega$, and we did not discuss the case of
$\omega$ being greater than the dynamical exponent, which is the regime
studied by Alava \etal. We agree that this regime is the most important 
one for SOC.

Alava \etal have chosen an observable that is independent of the
external drive for sufficiently large $\omega$,
but its finite-size scaling turns out to depend on the
scaling of the dissipation. This can be seen by a 
finite-size scaling analysis
of the characteristic avalanche size $s_c$.
According to Ref.~[\onlinecite{AlavaETAL:2008}], Eq.~(3),
$s_c\propto\xi_s^D$, where $\xi_s$ is some ``cut-off scale''. 
Usually,
the
exponent $D$ is reserved for the finite size scaling of
$s_c$, also known as the avalanche dimension \cite{ChristensenETAL:2004}, 
which we call $D_L$ in the following, so that $s_c\propto L^{D_L}$.
Combining Eq.~(4) and Eq.~(3) of \cite{AlavaETAL:2008}, they find
$D_L=\kappa/(2-\tau_s)$ with $\tau_s$ being
independent of external drive and dissipation. This is a very surprising
result, because 
regardless of whether or not the AS-mechanism applies,
the avalanche dimension $D_L$ is deeply rooted in the model and in general
directly related to the field theory of the corresponding depinning
transition
\cite{ChristensenETAL:2004,PaczuskiBoettcher:1996,Pruessner:2003,Alava:2004}.
There are several models
\cite{PanETAL:2005,Malthe-Sorenssen:1999,ChristensenETAL:2004}
which display a
universal avalanche dimension $D_L$ and an avalanche size exponent 
$\tau_s=2-\gamma_1/D_L$ 
which depends on 
the details of the driving of the model
\cite{ChristensenETAL:2004}, with the first
moment scaling like $\ave{s}\propto L^{\gamma_1}$. 
For these
conservative models, 
it is straight
forward to devise a method to produce any exponent $\tau_s$ in the
interval $[2-2/D_L,2-1/D_L]$ by effectively tuning $\gamma_1$. This can
be achieved by 
changing the driving mechanism,
\footnote{As long as the model is
conservative, individual particles effectively perform a random walk and
the first moment of the avalanche size is a function of the spatial
distribution of
the driving. For example, if a one-dimensional model is driven at site $x_0$ with
two open boundaries, then
the first moment is $(L+1-x_0) x_0$ where $L$ is the system
size. By changing $x_0$ as a power law of $L$, the exponent $\gamma_1$
changes accordingly. For example, $x_0=\sqrt{L}$, produces
$\gamma_1=3/2$.} 
as the driving mechanism leaves $D_L$ unchanged.

We would have
expected that changing $\gamma_1$ by introducing dissipation would have
the same effect, \ie varying $\tau_s$ and constant $D_L$. 
However, Alava \etal find that $D_L$ depends on $\kappa$,
while $\tau_s$ remains unchanged, a very interesting 
numerical finding we do not dispute.
It has thus been established that, under the appropriate conditions, 
both $\tau_s$ and $D_L$ can be tuned. Importantly, $D_L$ can be tuned
in the SOC-regime of the AS approach (large $\omega$) by changing $\kappa$.

Inasmuch as avalanche exponents pertain to the discussion the universality of
$\tau_s$ within the AS approach supports the case for universal SOC being
generated by the AS mechanism. But, the implicit finding by Alava
\etal that $D_L$ depends on $\kappa$ confirms that, apparently, the AS approach
does not produce universal SOC.

In order to address the question whether the AS-mechanism is operating
in SOC models (universal or not), one needs to probe its presence either directly or test
its implications. In
\cite{PetersPruessner:2006} we have shown that the AS-mechanism 
would (almost always) 
imply a vanishing
relative correlation length $\xi/L$ and a finite-size scaling of 
the AS order parameter, characterized by exponents 
$\beta/\mu$ and $\gamma/\mu$, 
that would depend on the scaling of dissipation
and drive, parameterized by $\kappa$ and $\omega$. 
We stated explicitly how $\beta/\mu$ and $\gamma/\mu$ depend on
$\kappa$ and $\omega$, while further analysis is necessary for the
dependence of the avalanche exponents on $\kappa$ and $\omega$.
At the present stage, a 
more promising route than studying avalanches therefore 
seems to be the study of AS observables in SOC systems.

\bibliography{articles,books}

\end{document}